\documentclass[11pt,twoside]{article}

\usepackage{baaa}

\usepackage{graphicx}
\usepackage{subfigure}
\usepackage{amssymb}
\usepackage[spanish,activeacute,english]{babel}
\usepackage[latin1]{inputenc}
\usepackage{verbatim}
\usepackage{amsmath}
\usepackage{amsfonts}
\usepackage{amssymb}
\usepackage{natbib}

\begin{document}
\myselectenglish
\vskip 1.0cm \markboth{Ocampo M. M. et al.}%
{Explorando el impacto de los gradientes qu\'imicos en los
procesos de mezcla del interior estelar}

\pagestyle{myheadings}

\vspace*{0.5cm}

\noindent PRESENTACI\'{O}N ORAL

\vskip 0.3cm
\title{Explorando el impacto de los gradientes qu\'imicos en
los procesos de mezcla del interior estelar}

\author{M.M. Ocampo$^{1,2}$, M.M. Miller Bertolami$^{1,2}$, L.G. Althaus$^{1,2}$ \& F.C. Wachlin$^{2}$}

\affil{%
(1) Instituto de Astrof\'{\i}sica de La Plata, CONICET-UNLP, Argentina \\
(2) Facultad de Ciencias Astron\'omicas y Geof\'{\i}sicas, Universidad Nacional de La Plata, Argentina \\
}

\begin{abstract}
During the various steps of stellar evolution are formed convectives zones that alter the chemical stratification in stars. Usually, in astrophysics is used the Mixing Length Theory (MLT) for modeling the convective movement and, in general, it is used with the Schwarzschild instability criterion, which neglects the impact of chemical composition gradients in the development of convection. However, towards the end of central helium burning and during the thermal pulses in the Asymptotic Giant Branch (AGB) are produced stratification processes with inversions in the chemical gradient that would produce instabilities beyond the ones predicted by the Schwarzschild criterion. These instabilities would alter the chemical profile in the white dwarfs, with respect to the one predicted by MLT, having observable consequences in the pulsational modes of such objects. In the present work we will explore an extension of MLT in which we will consider the chemical instabilities as generators of convectives and non-convectives instabilities. This theory will be applied in stellar evolution models in comparison with standard MLT and a double diffusive mixing theory, discussing the benefits and shortcomings of each one.
\end{abstract}

\begin{resumen}
Durante las diferentes etapas de evoluci\'on estelar se forman diversas zonas convectivas que alteran la estratificaci\'on qu\'imica de las estrellas. Usualmente, en astrof\'isica se utiliza la denominada teor\'ia de la longitud de mezcla (MLT) para tratar el transporte de calor y, normalmente, se utiliza en conjunto con el criterio de inestabilidad de Schwarzschild, el cual desprecia el impacto de los gradientes de composici\'on qu\'imica en el desarrollo de la convecci\'on. Sin embargo, hacia el final de la quema del helio en el n\'ucleo y durante los pulsos t\'ermicos en la Rama Asint\'otica de las Gigantes, se producen procesos de estratificaci\'on en los cuales ocurren inversiones del gradiente qu\'imico, los cuales producir\'ian inestablabilidades adicionales a las predichas por el criterio de Schwarzschild. Estas inestabilidades  alterar\'ian el perfil qu\'imico resultante en las estrellas enanas blancas y pre-enanas blancas respecto del predicho por MLT, teniendo consecuencias observables en los modos de pulsaci\'on de dichos objetos. En el presente trabajo exploraremos una extensi\'on de MLT en la cual consideraremos las inestabilidades qu\'imicas como generadoras de inestabilidades convectivas y no convectivas. Esta teor\'ia ser\'a aplicada en modelos de evoluci\'on estelar, en conjunto con MLT est\'andar y con una tercera teor\'ia de longitud de mezcla doblemente difusiva, y compararemos los resultados obtenidos, discutiendo los beneficios y dificultades de cada una.
\end{resumen}

\section{Introducci\'on}
\label{S_intro}

Las estrellas constituyen los pilares sobre los cuales el Universo est\'a construido y, como tal, su
estudio ha ganado inter\'es a lo largo de los a\~nos. En particular, las estrellas enanas blancas,
objetos longevos y extremadamente densos del tama\~no de la Tierra, no son la excepci\'on. Al constituir el estado evolutivo final para m\'as del 95\% de todas las
estrellas, juegan un rol de gran importancia en nuestro entendimiento de la formaci\'on
y evoluci\'on estelar, la evoluci\'on de sistemas planetarios y la historia de nuestra Galaxia misma.
El estudio de las enanas blancas por lo tanto resulta de relevancia central en una amplia variedad
de t\'opicos de la astrof\'isica moderna (Althaus et al., 2010; Farihi, 2016; Salaris \&
Cassisi, 2018).

La estructura qu\'imica de una enana blanca est\'a directamente relacionada al canal evolutivo que
llev\'o a su formaci\'on (C\'orsico et al., 2019). Por esto, mejoras en el modelado de estrellas enanas
blancas requieren de secuencias evolutivas basadas en una descripci\'on actualizada de los
procesos responsables de formar y evolucionar la estructura qu\'imica tanto antes y durante la
etapa de enana blanca. El estudio de estos procesos responsables de cambios qu\'imicos
internos y su impacto en la evoluci\'on de las enanas blancas son de fundamental importancia a la
hora de obtener determinaciones de edades de poblaciones estelares en base a sus enanas
blancas, tasas de acreci\'on de planetesimales, propiedades pulsaciones y magn\'eticas de las
enanas blancas y propiedades de part\'iculas elementales.

La mezcla de elementos en el interior estelar es, usualmente, tratada mediante la teor\'ia de longitud de mezcla (MLT, por sus siglas en ingl\'es). Sin embargo, esta teor\'ia no tiene en cuenta la convecci\'on inducida por los gradientes qu\'imicos negativos ni los procesos doblemente difusivos, como la mezcla termohalina y la semiconvecci\'on. Esta descripci\'on puede resultar insuficiente durante la evoluci\'on en la Rama Asint\'otica de las Gigantes (AGB, por sus siglas en ingl\'es) y, en particular, durante los pulsos t\'ermicos en los cuales se producen inversiones en el gradiente qu\'imico. Una teor\'ia que tiene en cuenta tanto las inestabilidades Rayleigh-Taylor como los procesos doblemente difusivos es, por ejemplo, la desarrollada en  Grossman et al. (1993); Grossman \&
Narayan (1993); y Grossman (1996) (GNA). Sin embargo, esta teor\'ia no considera el efecto de la degeneraci\'on electr\'onica en el medio, la cual se vuelve considerable en el interior profundo de las estrellas AGB y enanas blancas. A su vez, las ecuaciones que describen esta teor\'ia son significativamente m\'as complejas y num\'ericamente inestables que aquellas correspondientes a MLT, lo que genera problemas adicionales al incluirlas en un c\'odigo de evoluci\'on estelar en etapas ya de por s\'i inestables. Recientemente,  Fuentes et al. (2023) desarrollaron una versi\'on doblemente difusiva de la MLT que s\'i contempla el escenario de degeneraci\'on electr\'onica. Sin embargo, esta teor\'ia ha sido desarrollada en el marco de una mezcla de dos componentes y para casos en los cuales el flujo qu\'imico es impuesto por un agente externo a los movimientos convectivos. Por lo tanto, no es directamente aplicable a la evoluci\'on estelar en general.

El objetivo del presente trabajo es introducir una extensi\'on simple de MLT que contemple las inestabilidades Rayleigh-Taylor y la mezcla termohalina, con vistas a ser generalizada posteriormente a casos que contemplen materia degenerada.

\section{Teor\'ia de mezcla y estrellas AGB}

Como se mencion\'o en la Secci\'on \ref{S_intro}, en astrof\'isica se suele
usar la MLT junto con el llamado criterio de inestabili-
dad de Schwarzschild para el tratamiento de la convec-
ci\'on en los interiores estelares. A continuaci\'on repasa-
remos ambos conceptos y c\'omo se extienden a un caso
m\'as general, siguiendo Kippenhahn et al. (2013). Luego
exploraremos las ecuaciones que regir\'an la din\'amica de
las inestabilidades Rayleigh-Taylor en dicho enfoque.
En este trabajo nos concentraremos en estudiar el
impacto de estos procesos en la estructura qu\'imica de
las estrellas en la AGB.

\subsection{Criterio de Ledoux, MLT y un poco m\'as} \label{sec2.1}

Consideremos un elemento de materia que se desplaza de su posici\'on original. Si el entorno es din\'amicamente estable, el elemento de materia ser\'a forzado a volver a su posici\'on original. En cambio, si es inestable, la teor\'ia de longitud de mezcla nos dice que el elemento se desplazar\'a una longitud t\'ipica, denominada longitud de mezcla (\textit{mixing length}), y se disolver\'a, mezcl\'andose con un nuevo entorno.Un an\'alisis de estabilidad de los elementos de material en el interior de la estrella nos dice que la estratificaci\'on de temperaturas es estable frente a desplazamientos radiales si se cumple
\begin{equation}
    \nabla < \nabla_\text{e} + \frac{\varphi}{\delta}\nabla_\mu,
\end{equation}
donde
\begin{eqnarray}
    \nabla\equiv \bigg(\frac{d\ln T}{d\ln P}\bigg)_\text{s},   \nabla_\text{e}\equiv \bigg(\frac{d\ln T}{d\ln P}\bigg)_\text{e}, \nabla_\mu \equiv \bigg(\frac{d\ln \mu}{d\ln P}\bigg)_\text{s},
\end{eqnarray}
 los sub\'indices e y s indican elemento o entorno \textit{(surrounding)}, respectivamente, y
\begin{equation}
    \varphi \equiv \bigg (\frac{\partial\ln \rho}{\partial \ln \mu}\bigg)\bigg|_{T,P} , \ \ \delta\equiv -\bigg(\frac{\partial\ln\rho}{\partial\ln T}\bigg)\bigg|_{P,\mu},
\end{equation}
siendo $T$ la temperatura, $P$ la presi\'on, $\rho$ la densidad y $\mu$ el peso molecular.
Al suponer que el elemento se desplaza de forma adiab\'atica tenemos que $\nabla_\text{e}=\nabla_\text{ad}$ y el critero de inestabilidad resulta
\begin{equation}
    \nabla_\text{rad} > \nabla_\text{ad}  + \frac{\varphi}{\delta}\nabla_\mu,
\end{equation}
\begin{figure}[!t]
\centering
\includegraphics[scale=0.5]{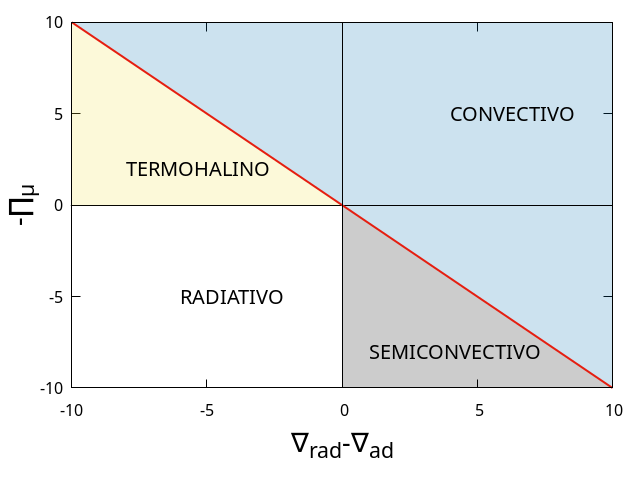}
\caption{Criterio de Ledoux (en rojo, el l\'imite) y las diferentes regiones de inestabilidad, donde $\Pi_\mu\equiv\frac{\varphi}{\delta}\nabla_\mu$. Observamos regiones inestables extra cuando $\nabla_\text{rad}<\nabla_\text{ad}$ y $\Pi_\mu<0$. Por encima del l\'imite se encuentran las inestabilidades convectivas de Rayleigh-Taylor, por debajo la difusi\'on t\'ermica da lugar a la mezcla termohalina. }
\label{FigLedoux}
\end{figure}
Esta expresi\'on se conoce como el \textit{criterio de Ledoux}. Despreciando el efecto de los gradientes qu\'imicos se obtiene el \textit{criterio de Schwarzschild}, normalmente usado en conjunto con la MLT,
\begin{equation}
    \nabla_\text{rad}>\nabla_\text{ad} .
\end{equation}
Al usar el criterio de Ledoux, en cambio, aparece una nueva regi\'on de inestabilidad convectiva, debida al impacto de los gradientes qu\'imicos (ver Figura \ref{FigLedoux}).\footnote{Al mismo tiempo desaparece una regi\'on convectiva que, al tener en cuenta la doble difusi\'on, es reemplazada por la semiconvecci\'on. En este trabajo la semiconvecci\'on no ser\'a considerada en MLT\#, tomando dichas regiones como convectivas usando MLT.} Esta nueva regi\'on convectiva corresponde a un tipo de inestabilidad de Rayleigh-Taylor (Garaud, 2021).

Adicionalmente, debemos considerar la mezcla doblemente difusiva conocida como termohalina. La combinaci\'on de las inestabilidades Rayleigh-Taylor junto con la mezcla termohalina, tiene como consecuencia que, independientemente del criterio de Schwarzschild usado en MLT, si $\nabla_\mu<0$ la regi\'on ser\'a inestable y deber\'a producirse alg\'un tipo de mezcla, din\'amica o doble difusiva seg\'un corresponda, como puede observarse en la Figura \ref{FigLedoux}.

Nosotros exploraremos una extensi\'on simple de MLT para tener en consideraci\'on las inestabilidades de Rayleigh-Taylor y la mezcla termohalina. En este trabajo indicaremos como MLT\# a esta extensi\'on de la MLT que incorpora el efecto de los gradientes qu\'imicos, en conjunto con la utilizaci\'on del criterio de Ledoux. Considerando la presencia de los gradientes qu\'imicos, y haciendo un an\'alisis similar al presentado por Kippenhahn et al. (2013) para MLT est\'andar, es posible obtener las siguientes ecuaciones para los movimientos convectivos:
\begin{equation}
    \nabla_\text{e}-\nabla_\text{ad}=2U\frac{\nabla-\nabla_\text{e}}{\sqrt{\nabla-\nabla_\text{e}-\frac{\varphi}{\delta}\nabla_\mu}},
\end{equation}
y
\begin{equation}
    \nabla-\nabla_\text{e}=\frac{8}{9}U\frac{\nabla_\text{rad}-\nabla}{\sqrt{\nabla-\nabla_\text{e}-\frac{\varphi}{\delta}\nabla_\mu}},
\end{equation}
donde
\begin{equation}
    U\equiv\frac{3acT^3}{c_P\rho^2\kappa l_m^2}\sqrt{\frac{8H_P}{g\delta}},
\end{equation}
siendo $a$ la constante de radiaci\'on, $c$ la velocidad de la luz, $c_P$ el calor espec\'ifico a presi\'on constante, $\kappa$ la opacidad y $l_m$ la longitud de mezcla de la teor\'ia. Por su parte $H_P\equiv-dr/d\ln P$ se denomina escala de altura de la presi\'on.

Con estas cantidades es posible obtener la velocidad de mezcla
\begin{equation}
    v^2=g\delta(\nabla-\nabla_\text{e}-\frac{\varphi}{\delta}\nabla_\mu)\frac{l_m^2}{8H_P} .
\end{equation}

Como ingrediente final para nuestro modelo es importante considerar la difusi\'on t\'ermica que da lugar a la mezcla termohalina. Al ser un proceso no din\'amico, es una mezcla mucho m\'as lenta que la inestabilidad convectiva. A pesar de eso, las escalas de tiempo de la mezcla son suficientemente r\'apidas para ser de importancia en la estrucura estelar, como veremos m\'as adelante. 
Este proceso se puede parametrizar de manera simple con la siguiente velocidad de mezcla  (Kippenhahn et al., 1980)

\begin{equation}
    v_T=\frac{4acT^3}{c_P\rho^2\kappa}\frac{\frac{\varphi}{\delta}\nabla_\mu}{(\nabla-\nabla_\text{ad})} .
\end{equation}

\subsection{Estrellas AGB y pulsos t\'ermicos}

Durante la evoluci\'on estelar, al final de la quema de helio en el n\'ucleo, las estrellas con masas iniciales $(M_i<8-10M_\odot)$ se ubican en la AGB en el diagrama de Hertzsprung-Russell. Estas estrellas poseen un n\'ucleo inerte de carbono y ox\'igeno rodeado por dos capas conc\'entricas donde se van quemando el helio y el hidr\'ogeno. A medida que estas capas van
avanzando hacia el exterior de la estrella, la capa que quema helio se volver\'a inestable, dando lugar a los denominados pulsos t\'ermicos.

Durante un pulso t\'ermico, la capa que quema helio se vuelve muy delgada e inestable, produciendo un embalamiento t\'ermico en la misma (Kippenhahn et. al 2013). Debido a la sensibilidad respecto a la temperatura de las reacciones nucleares de la quema del helio, un peque\~no incremento en la misma produce un gran aumento en la tasa de la reacci\'on. La luminosidad de helio $L_\text{He}$ se incrementa en varios \'ordenes de magnitud, sin embargo, sin aumentar la luminosidad total de la estrella. Esto es debido a que la mayor\'ia de la energ\'ia liberada expande las capas superiores a la de la quema del helio, incluyendo a la capa que quema hidr\'ogeno. Esta expansi\'on enfr\'ia la capa de hidr\'ogeno, reduciendo considerablemente su luminosidad $L_\text{H}$. Al haberse expandido, la capa de quema de helio no es m\'as inestable, y toda la regi\'on comienza a contraerse nuevamente, encendiendo nuevamente la quema de hidr\'ogeno. Luego de contraerse lo suficiente, todo el sistema recupera asint\'oticamente su estado original. Luego de un per\'iodo de estabilidad de miles o decenas de miles de a\~nos, la capa que quema helio vuelve a tornarse inestable 
y ocurre el siguiente pulso t\'ermico. Durante los flashes de helio se forma una zona convectiva intermedia, entre el n\'ucleo de carbono-ox\'igeno y la envoltura convectiva. Esta zona draga elementos del n\'ucleo hacia capas superiores. Normalmente, la intensidad de los flashes de helio ir\'a aumentando en los sucesivos pulsos t\'ermicos. Cuando esta intensidad sea lo suficientemente alta, la capa de hidr\'ogeno se apaga por completo y la envoltura convectiva es capaz de penetrar en las zonas ricas en carbono y ox\'igeno, drag\'andolos y enriqueci\'endose de metales en lo que se conoce como tercer \textit{dredge-up}.

Cada uno de los flashes de helio y la formaci\'on de la zona convectiva intermedia dejan como remanente estratificaciones qu\'imicas muy marcadas, con forma de ``picos'' en los perfiles de carbono-ox\'igeno. Esto se traduce en inversiones del gradiente qu\'imico que, considerando el criterio de Ledoux y la doble difusi\'on, deber\'ian ser inestables. Sin embargo, en esta regi\'on, $\nabla_\text{rad}<\nabla_\text{ad}$, por lo que las mismas son estables seg\'un el criterio de Schwarzschild y, por ende, cuando los c\'alculos se realizan con la MLT est\'andar, las mismas no se mezclan.
En la Secci\'on \ref{sec:simulaciones} efectuaremos simulaciones para comparar los perfiles qu\'imicos resultantes considerando MLT, MLT\# y GNA.

\section{Simulaciones}\label{sec:simulaciones}

Los c\'alculos de las secuencias fueron efectuados usando \texttt{LPCODE} (Miller Bertolami, 2016; Althaus et al., 2020). Presentamos en este art\'iculo los resultados obtenidos para estrellas de  $1M_\odot$, $1.5M_\odot$ y $3M_\odot$. Con el objetivo de eliminar incertezas num\'ericas, y sabiendo que el inter\'es del presente trabajo se enfoca en comparar los efectos de las diferentes teor\'ias de mezcla en la AGB, toda la secuencia principal y quema central del helio es calculada de la misma manera, usando MLT junto al criterio de Schwarzschild.

En todas las secuencias se consider\'o una metalicidad inicial de Z=0.015 y longitud de mezcla $l_m=1.822$, mientras que el resto de la f\'isica considerada coincide con la descrita en la Secci\'on 2 de  Miller Bertolami, 2016. Esta elecci\'on permite replicar varios observables de la secuencia principal hasta la fase de enana blanca.

Durante cada pulso t\'ermico se forma una discontinuidad en el perfil qu\'imico de la estrella, que se puede observar en forma de ``picos'' en la Figura \ref{Figura}. Sin embargo, puede permaneceer un remanente de helio que estabilice la regi\'on. Una vez que este remanente de helio se queme, el gradiente qu\'imico se invertir\'a. Al usar MLT\# y GNA esa regi\'on se mezclar\'a y los picos se suavizan {en los siguientes pulsos.  MLT, por su parte, no mezcla estas regiones ya que las mismas no son inestables seg\'un el criterio de Schwarzschild. A su vez, el efecto de MLT\# y GNA se ve suprimido en la secuencia de $1M_\odot$, al no quemarse el remanente de helio luego de los cuatro pulsos que tuvo la simulaci\'on, resultando en una zona m\'as estable.  En el caso de la secuencia de $3M_\odot$ puede notarse una diferencia adicional en el caso del modelo calculado con la GNA. El mismo muestra un n\'ucleo un poco m\'as grande, que se manifiesta en un corrimiento hacia la derecha de los perfile qu\'imicos mostrados en la figura  \ref{Figura}. Esto se debe a que la GNA incluye el tratamiento de las regiones semiconvectivas, lo que aten\'ua el efecto del tercer \textit{dredge-up}, dando lugar a un mayor crecimiento del n\'ucleo.

\begin{figure}[!t]
\centering
\includegraphics[scale=0.35]{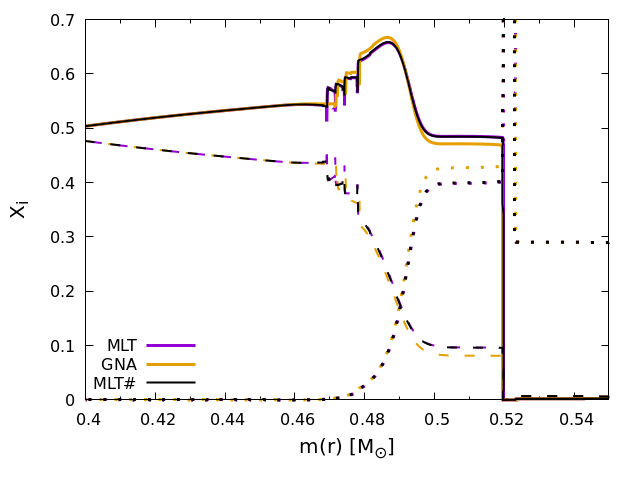}
\includegraphics[scale=0.35]{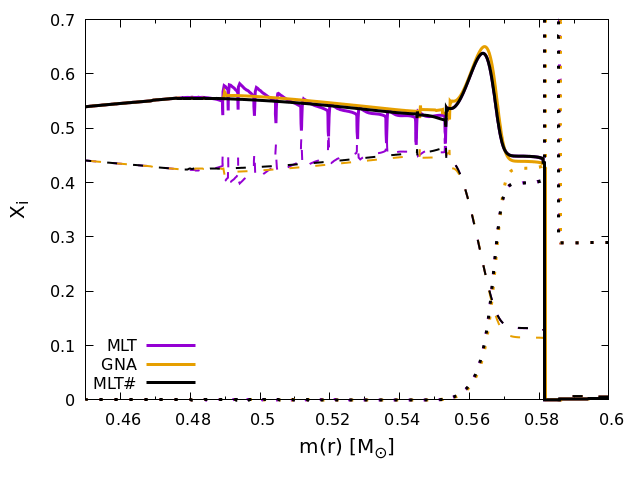}
\includegraphics[scale=0.35]{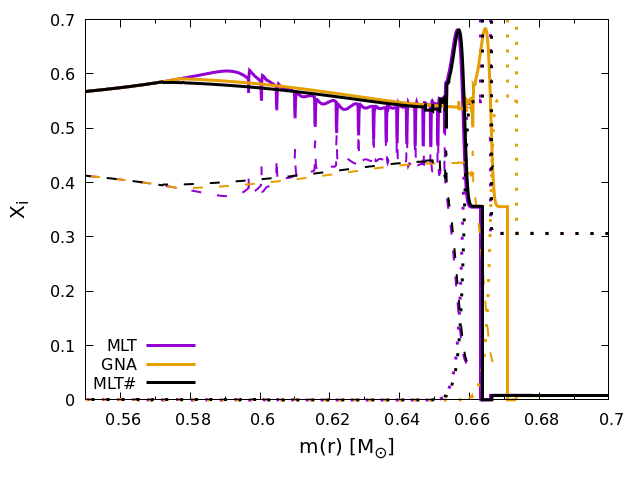}
\caption{Abundancias qu\'imicas $X_i$ de estrellas de 1$M_\odot$ (panel superior), 1.5$M_\odot$ (medio) y $3M_\odot$ (inferior) en la etapa final de la AGB, usando MLT, MLT\# y GNA, donde las l\'ineas s\'olidas representan la abundancia de carbono, las l\'ineas a trazos la abundancia de ox\'igeno y las punteadas indican helio. Observamos que MLT\# y GNA producen el mismo efecto de mezcla y suavizado en los perfiles qu\'imicos. En el caso de 1$M_\odot$ la mezcla es menor.} 
\label{Figura}
\end{figure}

\section{Conclusiones}

En el presente trabajo se analizaron las inestabilidades producidas por el impacto de los gradientes qu\'imicos en los interiores estelares. Para el caso de gradiente qu\'imico negativo siempre se tiene alg\'un tipo de inestabilidad, ya sea din\'amica o doble difusiva (Figura \ref{FigLedoux}). Exploramos una forma simple de extender MLT para que la misma contemple el escenario de las inestabilidades de Rayleigh-Taylor. Esto, en combinaci\'on con la mezcla termohalina fue introducido en el c\'odigo de evoluci\'on estelar \texttt{LPCODE} bajo el nombre de MLT\# para analizar la evoluci\'on de los perfiles qu\'imicos en los interiores de las estrellas en la AGB. Paralelamente, se corrieron simulaciones usando el equema est\'andar de MLT y la teor\'ia doble difusiva denominada GNA. Observamos (Figura \ref{Figura}) que, al introducir las inestabilidades qu\'imicas, el perfil resultante es diferente al predicho por MLT, encontr\'andose suavizado al haberse mezclado los picos formados en cada pulso t\'ermico. A futuro se analizar\'a el impacto de estas diferencias en los modos normales de oscilaci\'on de las estrellas GW Vir  (C\'orsico et al., 2019).

\references

\texttt{Althaus L.G., et al., 2010, A\&A Rv, 18, 471}

\texttt{Althaus L.G., et al., 2020, A\&A, 633, A20}

\texttt{C\'orsico A.H., et al., 2019, A\&A Rv, 27, 7}

\texttt{Farihi J., 2016, NewAR, 71, 9}

\texttt{Fuentes J.R., et al., 2023, ApJ, 950, 73}

\texttt{Garaud P., 2021, arXiv e-prints, arXiv:2103.08072}

\texttt{Grossman S.A., 1996, MNRAS, 279, 305}

\texttt{Grossman S.A., Narayan R., 1993, ApJS, 89, 361}

\texttt{Grossman S.A., Narayan R., Arnett D., 1993, ApJ, 407, 284}

\texttt{Kippenhahn R., Ruschenplatt G., Thomas H.C., 1980, A\&A,
91, 175}

\texttt{Kippenhahn R., Weigert A., Weiss A., 2013, Stellar Structure
and Evolution}

\texttt{Miller Bertolami M.M., 2016, A\&A, 588, A25}

\texttt{Salaris M., Cassisi S., 2018, PhyS, 93, 044002}

\end{document}